\documentclass[prb,superscriptaddress,twocolumn]{revtex4}
\usepackage{epsfig}

\begin{document}
\title{Spatial ordering of charge and spin in quasi one-dimensional
Wigner molecules}
\author{B. Szafran}
\affiliation{Departement Natuurkunde, Universiteit Antwerpen
(Campus Drie Eiken), B-2610 Antwerpen, Belgium}
\affiliation{Faculty of Physics and Nuclear Techniques, AGH
University of Science and Technology, al. Mickiewicza 30, 30-059
Krak\'ow, Poland}
\author{F.M. Peeters}
\affiliation{Departement Natuurkunde, Universiteit Antwerpen
(Campus Drie Eiken), B-2610 Antwerpen, Belgium}
 \author{S. Bednarek}
 \affiliation{Faculty of Physics and Nuclear Techniques, AGH
University of Science and Technology, al. Mickiewicza 30, 30-059
Krak\'ow, Poland}
 \author{T. Chwiej}
 \affiliation{Departement Natuurkunde, Universiteit Antwerpen
(Campus Drie Eiken), B-2610 Antwerpen, Belgium}
 \affiliation{Faculty of Physics and Nuclear Techniques, AGH
University of Science and Technology, al. Mickiewicza 30, 30-059
Krak\'ow, Poland}
 \author{J. Adamowski}
\affiliation{Faculty of Physics and Nuclear Techniques, AGH
University of Science and Technology, al. Mickiewicza 30, 30-059
Krak\'ow, Poland}

\begin{abstract}
Few-electron systems confined in quasi one-dimensional quantum
dots are studied by the configuration interaction approach. We
consider the parity symmetry of states forming Wigner molecules in
large quantum dots and find that for the spin-polarized Wigner
molecules it strictly depends on the number of electrons. We
investigate the spatial spin-ordering in the inner coordinates of
the quantum system and conclude that for small dots it has a
short-range character and results mainly from the Pauli exclusion
principle while the Wigner crystallization in large dots is
accompanied by spin ordering over the entire length of the dot.
\end{abstract}

 \maketitle
\section{Introduction}
Strong confinement of charge carriers in two directions results in
reduction of their degrees of freedom to a single one, i.e., in
quasi one-dimensional motion. Such one-dimensional systems are
realized typically in split-gate\cite{tarucha,PSP} and
cleaved-edge overgrowth\cite{yacoby} semiconductor quantum wires,
as well as in carbon nanotubes,\cite{CN} but can also be realized
in finite-size systems, i.e., in anisotropic quantum
dots\cite{hawrylak} or quantum rings.\cite{MKOSKINEN} There is a
renewed interest in the one-dimensional systems related to the
recent progress of vapour-liquid-solid fabrication of quantum
wires of very high quality. \cite{vl1,vl2,vl3}

The present paper is devoted to electron systems confined in
one-dimensional quantum dots and in particular to their Wigner
crystallization\cite{Wigner} appearing when the electron-electron
interaction dominates over the kinetic energy. Wigner electron
solids (Wigner molecules) are predicted to appear in large
dots\cite{WM} or in strong magnetic fields.\cite{ReiMan} In the
Wigner molecules the charge density separates into distinct charge
maxima each corresponding to one of confined electrons. Formation
of Wigner molecules in the ground-state charge density in
one-dimensional quantum dots was previously obtained in exact
diagonalization\cite{JA,Hkramer,2e} and density functional
approach.\cite{Nieminen}  In one-dimensional dots the Wigner
localization appears in the laboratory frame, in contrast to the
inner-coordinate crystallization appearing in circular quantum
dots,\cite{ReiMan} including quantum rings. Transport properties
of Wigner crystals formed in open infinite one-dimensional systems
have also been studied.~\cite{Glazman,Schultz} The Luttinger
liquid formalism has been applied\cite{Anfuso} to quantum wires
with box-like boundary conditions, i.e., to the one-dimensional
quantum dots. Melting of classical one-dimensional Wigner crystals
has recently been described.\cite{Piacente}

We study the quasi one-dimensional quantum dots using a
configuration interaction approach with the effective
electron-electron interaction potential which we derived
recently.~\cite{efpot} This work is a generalization of our exact
two-electron study~\cite{2e} to larger number of electrons. In the
weak confinement limit the ground-state becomes nearly degenerate
with respect to the spin configuration of the electron
system.\cite{Hkramer,2e} Similar approximate degeneracy has been
found in quantum rings of large radius.\cite{qrk} In this paper we
study the parity symmetry of the nearly degenerate states forming
Wigner molecules in large dots. We show that for spin-polarized
electrons the Wigner localization is formed only for one (even or
odd) spatial parity of the state strictly dependent on the number
of electrons. We present this dependence in the form of a theorem
for which we provide a rigorous analytical proof. The found
dependence of the parity of one-dimensional Wigner molecule states
on the number of electrons is similar to the appearance of the
magic angular momenta states for which Wigner crystallization is
possible in circular dots.\cite{magic1,magic2} Furthermore, we
discuss an inhibition of Wigner crystallization by a perturbation
of the confinement potential through a central inversion-invariant
potential well.

Magnetic spin-ordering of electrons in one-dimensional space has
been extensively studied\cite{kolo} in Hubbard models which, in
one dimension with only nearest-neighbor hopping interactions,
predict the appearance of a low-spin ground state.\cite{tasaki}
This is a consequence\cite{tasaki} of the Lieb-Mattis
theorem\cite{LM} which implies that without spin-dependent
interactions the ground-state of one-dimensional electron systems
corresponds to the lowest possible spin quantum number ($S$=0 or
$1/2$). This feature generally does not have to result in any
spatial spin ordering. In this paper we use the exact numerical
solution of the Schr\"odinger equation to investigate the spatial
distribution of spins in the one-dimensional quantum dot and the
relation between the charge and spatial spin ordering in the
Wigner crystallization limit. We find that Wigner crystallization
is accompanied by a long-range spin-ordering in the inner
coordinates of the system instead of a spin-symmetry breaking
predicted by density functional theory.\cite{Nieminen,Reiman} In
the ground-state this ordering has a clear antiferromagnetic
character.

This paper is organized as follows. In Section II we present the
theoretical method. Section III contains the results for the
Wigner localization and ground state degeneracy of the few
electron systems. In Section IV we present proof for the
dependence of the parity of spin-polarized Wigner molecules on the
number of electrons. Section V contains discussion of the effect
of a central defect on Wigner crystallization. In Section VI the
study of spin ordering is presented. Section VII contains our
summary and conclusions.

\section{Theory}
We consider $N$ electrons confined in a quasi-one-dimensional
quantum dot with strong lateral harmonic-oscillator confinement
potential. The Hamiltonian of the system reads,
\begin{equation}
H=\sum_{i=1}^N h_i + \sum_{i=1}^N\sum_{j>i}^N \frac{\kappa}{
r_{ij}} , \label{H0N}
\end{equation}
where $h$ stands for the single-electron Hamiltonian,
\begin{equation}
h=-\frac{\hbar^2}{2m^*}\nabla^2 +\frac{m^*\omega^2}{2}(x^2+y^2)
+V(z), \label{H1}
\end{equation}
$V(z)$ is the confinement potential in the $z-$direction. For a
strong lateral harmonic-oscillator confinement energy ($\hbar
\omega$) the movement of electrons in the $(x,y)$ plane is frozen
to the harmonic-oscillator ground state. Then, one can perform
integration\cite{efpot} over the lateral degrees of freedom which
results in the following Hamiltonian,
\begin{eqnarray}
H=&N\hbar\omega+\sum_{i=1}^Nh_i^{1D} + \sum_{i=1}^N\sum_{j>i}^N
 \left(\pi/2\right)^\frac{1}{2}(\kappa/l) \nonumber \\ & \times \mathrm{erfc}\left(z_{ij}/2^{1/2}l\right) \exp(z_{ij}^2/2l^2) , \label{HN}
\end{eqnarray}
where $z_{ij}=|z_i-z_j|$ and \begin{equation}
h^{1D}=-\frac{\hbar^2}{2m^*}\frac{d^2}{d z^2} +V(z) \label{jeje}
\end{equation} is the single-electron one-dimensional Hamiltonian.
In the following we will neglect the first term in Eq. (\ref{HN}),
i.e., the lateral confinement energy which is independent of the
form of wave functions in the $z$ direction. The last term in Eq.
(\ref{HN}) is the effective interaction energy \cite{efpot} for
electrons in a quasi-one-dimensional environment resulting from
integration of the Coulomb potential over the lateral coordinates,
$m^*$ is the effective mass,
$\kappa=e^2/4\pi\varepsilon_0\varepsilon$, $\varepsilon$ is the
dielectric constant,  and $l=\sqrt{\hbar/m^*\omega}$. We assume
$V(z)=V_{well}(z)$, a rectangular potential well of depth
$V_0=200$ meV and width $d$. We adopt GaAs material parameters,
i.e., $m^*=0.067$ $m_{e0}$, $\epsilon=12.4$ as well as $\hbar
\omega = 10$ meV ($l=10.66$ nm) for the lateral confinement
energy. Calculations have been performed for $N=2,\dots,5$
electrons by the configuration interaction approach with a basis
set of Slater determinants built with single-electron
spin-orbitals. Spatial single-electron wave functions have been
obtained by numerical diagonalization of the finite-difference
version of the single-electron one-dimensional Hamiltonian
(\ref{jeje}) on a mesh of points. In construction of the Slater
determinants with required spin and parity symmetries we use the
spatial wave functions of up to 8 lowest-energy single-electron
states which results in a Slater determinant basis size of up to
1520 elements and an accuracy better than 0.01 meV.

\begin{figure*}[htbp]{\epsfxsize=170mm
                \epsfbox[24 638 571 775]{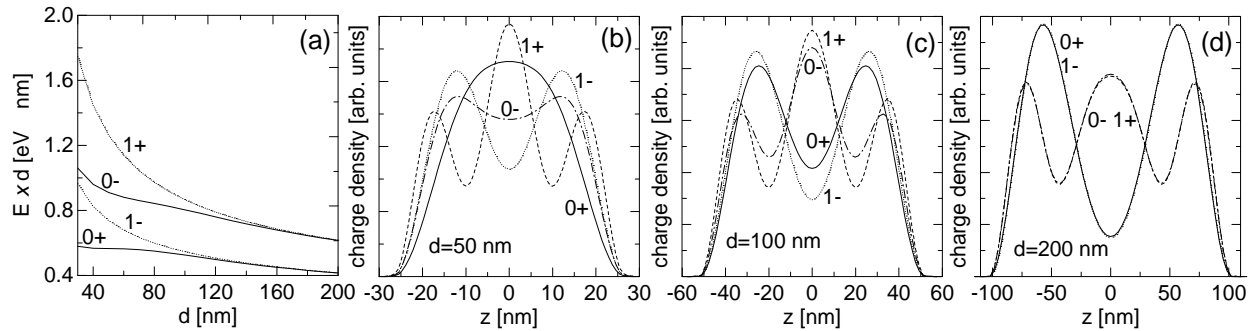}}\newline
\caption{(a) Lowest energy levels multiplied by the dot length for
$N=2$. Numbers close to the curves denote the total spin quantum
number of the corresponding states and signs $+$, $-$ stand for
even and odd parity symmetry. (b), (c), (d) - charge density of
$0+$, $1-$, $1+$ and $0-$ states plotted with solid, dotted,
dashed and dash-dotted lines for $d=50$, 100 and 200 nm,
respectively.
 } \label{2e}
\end{figure*}

\begin{figure*}[htbp]{\epsfxsize=170mm
                \epsfbox[30 603 590 747]{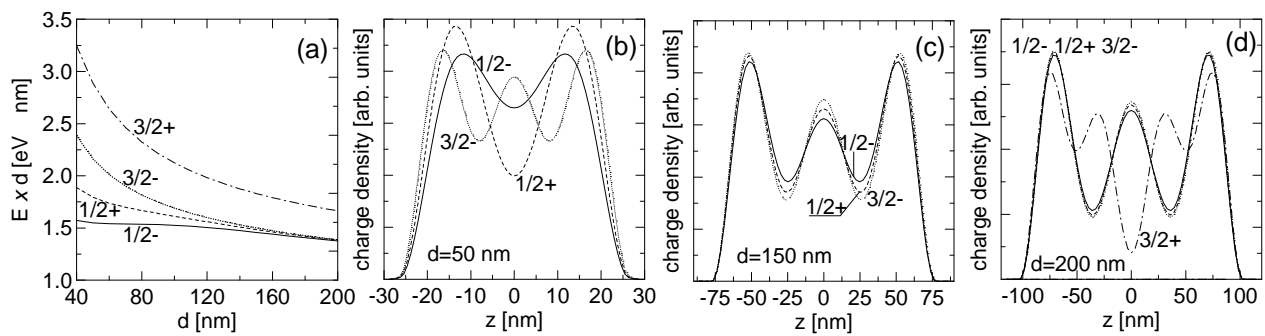}}\newline
\caption{(a) Lowest energy levels multiplied by the dot length for
$N=3$. (b), (c), (d) - charge density of $1/2-$, $1/2+$ and $3/2-$
states plotted with solid, dashed and dotted lines for $d=50$, 150
and 200 nm, respectively. In (d) the charge density of the $3/2+$
state is shown by the dash-dotted curve. \label{3e}
 }
\end{figure*}

The present approach is based on the assumption that only the
lowest state of the lateral ($x,y$) quantization is occupied. We
performed test calculations for 2, 3 and 4 electrons to check the
validity of this approach. We allowed the electrons to occupy also
the $p$-type lowest excited state of the lateral quantization with
angular momentum $\pm \hbar$. Inclusion of $p$ states not only
allows for determination of the critical well length above which
the $p$-shell is emptied, but it is also helpful to estimate the
importance of the angular correlations in the $x-y$ plane. The
Coulomb matrix elements were evaluated using effective interaction
potentials derived with the use of the Fourier transform
technique.\cite{efpot}  We have obtained the following results:
the $p$ shell is left empty for $d>39$ and 41 nm for $N=3$ and 4,
respectively (for 2 electrons the $p$ shell is never occupied).
Accounting for the $x-y$ correlations via inclusion of the
$p$-type orbitals in the configuration interaction basis lowers
the 2-electron total energy estimates by 0.18, 0.12, 0.01 and
$10^{-4}$ meV for $d=40, 50, 100$ and $200$ nm respectively. This
"lateral correlation energy" value for the same values of $d$ are
equal to 0.4, 0.3, 0.08 and $4\times 10^{-3}$ meV for $N=3$ and
1.18, 0.67, 0.23 and 0.03 for $N=4$, respectively. The energy
overestimation in the range of $d$ studied further is never
significant and the present approach is nearly exact in the Wigner
localization regime.

\section{Ground state degeneracy and Wigner crystallization}

In this paper we label the states by their total spin $S$ and
parity quantum numbers using the notation: $S\pm$, where the
positive (negative) sign stands for even (odd) parity. We discuss
only the lowest-energy states for a given spin-orbital symmetry.
Fig. \ref{2e}(a) shows the lowest energy levels of the 2-electron
system multiplied by the dot length $d$ as functions of $d$. For
large dots the states $0+$ and $1-$ as well as $0-$ and $1+$
become mutually degenerate. For large values of $d$ potential
energy related to penetration of electrons into the barrier region
is negligible, the kinetic energy scales as $1/d^2$ and the
Coulomb energy as $1/d$. Therefore, the product of energy and dot
length for large $d$ behaves as $f(d)=C+D/d$ function, where the
constants $C$ and $D$ are related to the Coulomb and kinetic
energy, respectively. The energy levels of the degenerate pairs of
states tend to different constants in the infinite $d$ limit which
is apparently due to different values of the Coulomb interaction
in these pairs of states. The evolution of the charge density for
growing length of the dot is shown in Figs. 1 (b-d). For large
dots [cf. Fig. \ref{2e}(d)] the charge densities of the degenerate
pairs of states become identical. In the ground state the charge
density has two pronounced maxima which indicates the separation
of electron charges into two charge islands, i.e., the Wigner
crystallization. Fig. \ref{2e} shows that the singlet-triplet
degeneracy obtained previously \cite{2e} for the two-electron
ground-state appears also in the first excited state.

\begin{figure*}[htbp]{\epsfxsize=175mm
                \epsfbox[15 608 583 754]{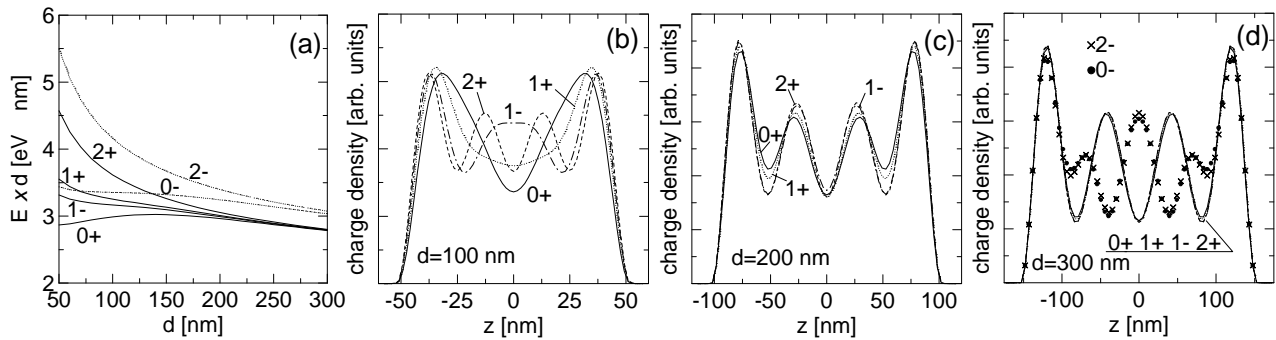}}\newline
\caption{(a) 4-electron energy levels multiplied by the dot
length. (b), (c), (d) - charge density of $0+$, $1-$, $1+$ and
$2+$ 4-electron states plotted with solid, dash-dotted, dotted and
dashed lines for $d=100$, 200 and 300 nm, respectively. In (d) the
charge densities of $2-$ and $0+$ states are marked with crosses
and dots, respectively.
 }\label{4e}
\end{figure*}

\begin{figure*}[htbp]{\epsfxsize=175mm
                \epsfbox[38 618 568 751]{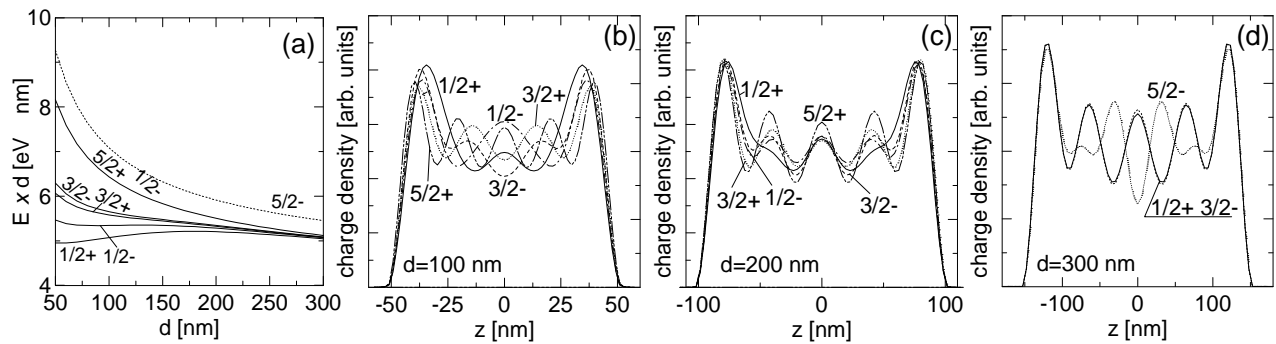}}\newline
\caption{(a) 5-electron energy levels multiplied by the dot
length. Even (odd) parity levels are plotted with solid (dotted)
lines. (b), (c) - charge density of $1/2+$, $1/2-$, $3/2+$, $3/2-$
and $5/2+$ states plotted with solid, dash-dotted, dotted, dashed
and dash-double-dot lines for $d=100$ and 200 nm, respectively. In
(d) the charge density of the $1/2+$, $3/2-$ and $5/2-$ state is
shown by solid, dashed and dotted lines, respectively [charge
densities of $1/2-$, $3/2+$ and $5/2+$ are almost identical with
the $1/2+$ and $3/2-$ charge densities are therefore omitted in
(d) for the sake of clarity].\label{5e}
 }
\end{figure*}

Fig. \ref{3e}(a) shows the energy levels and Figs. \ref{3e}(b-d)
the charge density for the lowest-energy states of the 3-electron
system for increasing $d$. For 3 electrons the Wigner molecule is
formed in states $1/2-$, $1/2+$ and $3/2-$ which become degenerate
for large $d$. In the state $3/2+$ the charge density exhibits
four maxima [cf. Fig. \ref{3e}(d)], which apparently prevents this
state to be degenerate with the ground state.

In the 4-electron system the ground state corresponds to $0+$
symmetry. The states $1-$, $1+$ and $2+$ for large dots [cf.
 \ref{4e}(a)] tend to the degeneracy with the ground state. The charge
 densities of these states for large
dots present four distinct maxima [cf. Fig. \ref{4e}(d)]. Energy
levels corresponding to states $0-$ and $2-$ are separated by a
significant energy distance from the ground state [cf. Fig.
\ref{4e}(a)] and in large dots they correspond to identical charge
densities with five maxima. The ground state charge density
evolution obtained for N=3 and 4 is in a qualitative agreement
with the results of Ref. [\cite{JA}].

Finally, in the 5-electron system the ground-state of $1/2+$
symmetry becomes degenerate with $1/2-,3/2+,3/2-$, and $5/2+$
states [cf. Fig. \ref{5e}(a)] forming Wigner molecules for large
dots [cf. Figs. \ref{5e}(b-d)]. The spin polarized state of odd
parity $5/2-$ does not become degenerate with the ground-state and
its charge density in large dots forms six maxima [cf. Fig
\ref{5e} (d)].

In the entire $d$ range and for all electron numbers studied the
order of the lowest energy levels for given total spin quantum
numbers (neglecting the parity) follow the order of the spin
quantum numbers, which is in agreement with the theorem  of Lieb
and Mattis.\cite{LM} In large dots a ground-state degeneracy
appears. In Ref. [\cite{Hkramer}] the degeneracy was interpreted
in terms of a vanishing tunnel coupling between the local minima
of the total $N$-dimensional potential energy. The present results
indicate that the nearly degenerate states possess the same charge
density in the laboratory frame. Moreover, we observe the
following regularities. In the limit of Wigner localization the
ground state of the $N$-electron system appears for $N$ different
pairs of the spin and parity quantum numbers.\cite{2doN} For even
electron numbers $N=2$ and $4$, $N$ charge maxima are formed only
for even parity states with $S=0$, while the odd parity zero-spin
states possess $N+1$ charge maxima [cf. Figs. 1(d) and 3(d)]. The
spin-polarized Wigner-localized state can only be formed for one
(even or odd) parity. Namely, the parity of the spin-polarized
Wigner molecule state is even for 4 and 5 electrons and odd for 2
and 3 electrons. The charge density of the spin-polarized state of
the other parity exhibits $N+1$ maxima, i.e., the state does not
form a Wigner molecule and as a consequence does not become
degenerate with the ground state even for large dots. This
conclusion will be cast into a theorem in the next section.

\section{Parity of spin-polarized Wigner molecule states}

Here we give an analytical proof of the theorem: {\it for an odd
number of electrons $N=2M+1$ as well as for an even number of
electrons $N=2M$ the parity of one-dimensional spin-polarized
Wigner-molecule state is even (odd) for even (odd) value of the
integer $M$}

\begin{figure}[htbp]{\epsfxsize=55mm
                \epsfbox[60 76 560 560]{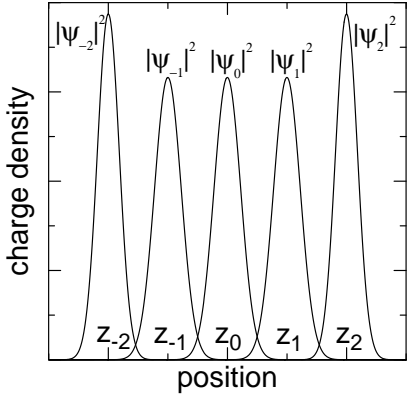}}\newline
\caption{ Illustration to the proof that for $N=2M+1$ or $N=2M$
electrons the parity of the spin-polarized state which exhibits
Wigner localization is accordant with the parity of $M$.
\label{proof}
 }
\end{figure}

We will present here the proof for an odd number of electrons (the
proof for even $N$ can be easily deduced from the present
demonstration). For odd $N$ one of the electrons resides near the
center of the dot (point $z_0=0$, cf. Fig. \ref{proof}), and the
others occupy spatially symmetric sites to the left and right of
the dot around points which satisfy $z_k=-z_{-k}$ for $k=\pm 1,\pm
2,\dots \pm M$. In the Wigner phase the total charge density
possesses $N$ maxima corresponding to the separate single-electron
charge densities. A single-electron charge $|\psi_k(z)|^2$ is
localized around point $z_k$. In the Wigner limit the overlap
between the single-electron charge densities vanishes (the proof
is only valid when this overlap is negligible), so the total
charge density can be expressed as their sum. Since the total
charge density is symmetric with respect to the origin the
following equality holds
\begin{equation} |\psi_k(-z)|^2=|\psi_{-k}(z)|^2, \end{equation}
which results in the following relation for the single-electron
wave functions
\begin{equation}
\psi_k(-z)=e^{i\phi_k}\psi_{-k}(z), \label{rel1}
\end{equation}
where the phase $\phi_k$ is a real number. Relation (\ref{rel1})
with changed sign of $k$ reads
\begin{equation}
\psi_{-k}(-z)=e^{i\phi_{-k}}\psi_{k}(z). \label{rel2}
\end{equation}
Phases $\phi_k$ and $\phi_{-k}$ are not independent. Changing the
sign of $z$ in Eq. (\ref{rel2}) and making use of relation
(\ref{rel1}) we arrive at
\begin{equation}
\psi_{-k}(z)=e^{i\phi_{-k}}\psi_{k}(-z)=e^{i(\phi_{-k}+\phi_{k})}\psi_{-k}(z),
\label{rel3}
\end{equation}
hence
\begin{equation}
\phi_{k}=-\phi_{-k}, \label{opphase}
\end{equation}
up to an unimportant multiple of $2\pi$. Considering relation
(\ref{rel1}) for $k=0$ and reminding that we arrive at the same
value $\psi_0(0)$ (non-zero for odd $N$) approaching the origin
from both positive and negative sides we arrive at $\phi_0=0$ and
consequently $\psi_0$ is an even parity function
\begin{equation} \psi_0(-z)=\psi_{0}(z). \end{equation}
Since the considered state is spin-polarized the spin and spatial
parts of the wave function can be separated into a product
\begin{eqnarray}
\chi(z_1,\sigma_1,\dots,z_N,\sigma_N)=\alpha(\sigma_1)\alpha(\sigma_2)\dots\alpha(\sigma_N)\times \nonumber \\
\times\Psi(z_1,z_2,\dots,z_N),
\end{eqnarray}
where $\alpha$ is an eigenfunction of the single-electron spin
$z$-component operator. The spatial wave function $\Psi$ can be
written as a Slater determinant\cite{1sd}
\begin{widetext}
\begin{equation}
\label{eq sslater} \Psi(z_1,z_2,\dots,z_N) =
 \left| \begin{array}{ccccc}
\psi_{-M}(z_1) & \psi_{-M+1}(z_1) & \dots \psi_{M-1}(z_1) &
\psi_{M}(z_1) \\
\psi_{-M}(z_2) & \psi_{-M+1}(z_2) & \dots \psi_{M-1}(z_2) &
\psi_{M}(z_2) \\ \dots & & & & \\
\psi_{-M}(z_N) & \psi_{-M+1}(z_N) & \dots \psi_{M-1}(z_N) &
\psi_{M}(z_N) \\
\end{array} \right|. \label{form8}
\end{equation}
We apply the parity operator on $\Psi$ and make use of properties
(6) and (9) obtaining
\begin{equation}
\label{eq slater} \Psi(-z_1,-z_2,\dots,-z_N) =
 \left| \begin{array}{ccccc}
e^{-i\phi_M}\psi_{M}(z_1) & e^{-i\phi_{M-1}}\psi_{M-1}(z_1) &
\dots e^{i\phi_{M-1}}\psi_{-M+1}(z_1) &
e^{i\phi_M} \psi_{-M}(z_1) \\
e^{-i\phi_M}\psi_{M}(z_2) & e^{-i\phi_{M-1}}\psi_{M-1}(z_2) &
\dots e^{i\phi_{M-1}}\psi_{-M+1}(z_2) &
e^{i\phi_M}\psi_{-M}(z_2) \\ \dots & & & & \\
e^{-i\phi_M}\psi_{M}(z_N) &e^{-i\phi_{M-1}} \psi_{M-1}(z_N) &
\dots e^{i\phi_{M-1}}\psi_{-M+1}(z_N) &
e^{i\phi_M}\psi_{-M}(z_N) \\
\end{array} \right|.
\end{equation}
Phase factors can be extracted from each of the determinant
columns, which yields
\begin{equation}
 \Psi(-z_1,-z_2,\dots,-z_N) =
e^{-i(\phi_M+\phi_{M-1}+\dots +\phi_{-M+1}+\phi_{-M})} \left|
\begin{array}{ccccc} \psi_{M}(z_1) & \psi_{M-1}(z_1) & \dots
\psi_{-M+1}(z_1) &
\psi_{-M}(z_1) \\
\psi_{M}(z_2) & \psi_{M-1}(z_2) & \dots \psi_{-M+1}(z_2) &
\psi_{-M}(z_2) \\ \dots & & & & \\
\psi_{M}(z_N) &\psi_{M-1}(z_N) & \dots \psi_{-M+1}(z_N) &
\psi_{-M}(z_N) \\
\end{array} \right|. \label{ostfo}
\end{equation}

\end{widetext}
The phases in front of the determinant in Eq. (\ref{ostfo}) cancel
according to property (\ref{opphase}). Exchanging $M$ pairs of
corresponding columns in the determinant we arrive at Eq.
(\ref{form8}) but multiplied by $(-1)^M$, which proofs that the
parity of spin-polarized one-dimensional Wigner molecule state is
determined by the odd or even value of $M$.

We have found that 2- and 4-electron zero-spin states can form a
Wigner-localized charge density only for even spatial parity. We
are unable to proof in general that the zero-spin state with
Wigner localization has to be of even parity for even $N$. But for
$N=2$ such a proof is easily given. In this case the spin and
spatial parts of the wave function can be separated as follows
\begin{eqnarray}
\chi^{0+}(z_1,\sigma_1,z_2,\sigma_2)
=\left[\alpha(\sigma_1)\beta(\sigma_2)-\alpha(\sigma_2)\beta(\sigma_1)\right]
\times
\nonumber \\
\times\left[\psi_1(z_1)\psi_{-1}(z_2)+\psi_{-1}(z_1)\psi_{1}(z_2)\right].
\label{w+}
\end{eqnarray}
Applying the parity operator to the spatial part of this wave
function and making use of the properties of the single-electron
wave functions given above we find that this wave function is of
even parity. Moreover, it follows that construction of a symmetric
spatial wave function for odd-parity singlet (zero-spin)
two-electron states ($0-$) requires at least three single-electron
functions, for instance, the function
\begin{eqnarray}
\Psi^{0-}(z_1,z_2)=
\psi_0(z_1)\psi_{1}(z_2)+\psi_1(z_1)\psi_{0}(z_2)\nonumber \\
-\psi_0(z_1)\psi_{-1}(z_2)-\psi_{-1}(z_{1})\psi_{0}(z_2),
\label{0-}
\end{eqnarray}
is of odd parity provided that we take zero phase shifts in
relation (\ref{rel1}).  Indeed, the $0-$ state for $N=2$ exhibits
three charge maxima [see Fig. 1(d)]. Moreover, construction of a
triplet antisymmetric spatial wave function with even parity
($1+$) also requires at least three localized functions, for
instance
\begin{eqnarray}
\Psi^{1+}(z_1,z_2)=
\psi_0(z_1)\psi_{1}(z_2)-\psi_1(z_1)\psi_{0}(z_2)\nonumber \\
+\psi_0(z_1)\psi_{-1}(z_2)-\psi_{-1}(z_{1})\psi_{0}(z_2),
\label{1+}
\end{eqnarray}
possess the required symmetries for zero phase shifts in relation
(\ref{rel1}). The charge density corresponding to wave functions
(\ref{0-}) and (\ref{1+}) is the same provided that the overlaps
between the functions $\psi_i$ are negligible. Fig. 1(d) shows
that the charge densities of the states $0-$ and $1+$ are indeed
indistinguishable. The area below the central maximum of the
probability density of degenerate 0- and 1+ states in Fig. 1(d) is
two times larger than the area below each of the extreme maxima,
which can be interpreted by saying that one of the electrons stays
in the neighborhood of the center of the system with $100\%$
probability while probabilities of finding the other one at the
left or right end of the well are equal to $50\%$. This feature is
in agreement with the probability amplitudes (\ref{0-}) and
(\ref{1+}). Although in the wave functions (\ref{0-}) and
(\ref{1+}) the electron positions are separated, this separation
has a non-classical character since the charge maxima at the left
and right ends of the dot correspond to sub-electron charges.
Therefore we do not refer to this separation as Wigner
localization. Average electron-electron distances in states
described by wave functions (\ref{0-}) and (\ref{1+}) are smaller
than in states $0+$, $1-$ with two charge maxima, which leads to a
larger value of the Coulomb interaction energy and consequently to
an energy separation between pairs of degenerate states $0+$,$1-$
and $0-,1+$ presented in Fig. \ref{2e}(a) in the weak confinement
limit.

\section{Wigner crystallization in the presence of a defect
potential} The presence of defects can significantly perturb the
Wigner crystallization in large systems. We consider here a thin
attractive cavity just deep enough to bind one electron. The
perturbed quantum dot potential is of the form
\begin{equation} V(z)=V_{well}(z)+V_{defect}(z), \label{cavity} \end{equation}
where $V_{defect}(z)=-50$  meV for $|z|<1$ nm and
$V_{defect}(z)=0$ for $|z|>1$ nm. The assumption that the defect
is localized in the center of the system does not perturb the
inversion invariance of the total potential.

\begin{figure}[htbp]{\epsfxsize=80mm
                \epsfbox[65 215 548 638]{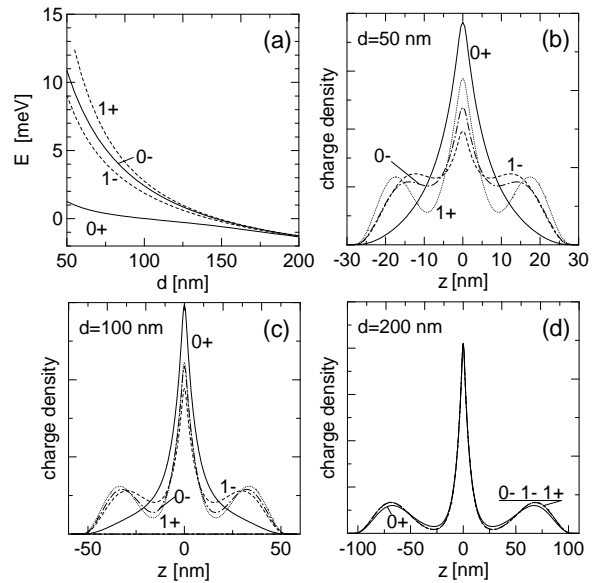}}\newline
\caption{(a) Lowest energy levels for $N=2$ as functions of the
length of the well with a central attractive cavity [Eq.
(\ref{cavity})]. Numbers close to the curves denote the total spin
quantum number of the corresponding states and signs $+$, $-$
stand for even and odd parity symmetry, respectively. (b), (c),
(d) - charge density of $0+$, $1-$, $1+$ and $0-$ states plotted
with solid, dotted, dashed and dash-dot curves for $d=50$, 100 and
200 nm, respectively.
 } \label{def2e}
\end{figure}

Fig. \ref{def2e}(a) shows that contrary to the unperturbed quantum
well potential [cf. Fig. \ref{2e}(a)] the $0-$ and $1+$ states
become degenerate with the $0+$ and $1-$ states.
 Figs. \ref{def2e}(b-d) show the evolution of the charge densities of
the four considered states with increasing size of the system. For
large well thickness [cf. Fig. \ref{2e}(d)] the charge densities
of these states become indistinguishable. One of the electrons is
trapped by the potential of the central cavity which results in
the sharp central peak of the charge density. The probability to
find the other electron at the left or right side of the origin
are equal. This differs essentially from the two-electron Wigner
molecule charge density in the unperturbed dot [cf. Fig.
\ref{2e}], for which the probability to find an electron in the
center of the well was negligible and for which each of the two
charge maxima could be associated with an integer electron charge.
The formation of three maxima in the charge density is possible
for all states [cf. Eqs (\ref{0-}) and (\ref{1+}), for $0-$ and
$1+$ states, similar formulas can be given for the other two].
Therefore, the ground-state tends to a fourfold degeneracy in
contrast to the double degeneracy for the unperturbed dot [cf.
Fig. 1(a)].

\begin{figure}[htbp]{\epsfxsize=80mm
                \epsfbox[65 148 560 630]{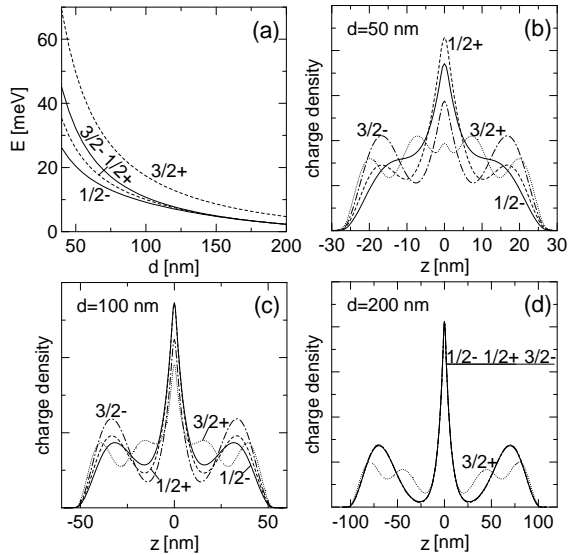}}\newline
\caption{(a) Lowest energy levels for $N=3$ as function of the
length of the well with a central attractive cavity [Eq.
(\ref{cavity})]. Numbers close to the curves denote the total spin
quantum number of the corresponding states and signs $+$, $-$
stand for even and odd parity symmetry. (b),(c),(d) - charge
density of $0+$, $1-$, $1+$ and $0-$ states plotted with solid,
dotted, dashed and dash-dot curves for $d=50$, 100 and 200 nm,
respectively.
 } \label{def3e}
\end{figure}

Fig. \ref{def3e} shows the lowest energy levels and corresponding
charge density evolution for the 3-electron system. Contrary to
the 2-electron system the central defect does not perturb the
number of charge maxima, Wigner localization appears similarly as
for the unperturbed dot [cf. Fig. \ref{3e}] for $1/2+$, $1/2-$,
and $3/2-$ states which become degenerate in the Wigner
localization limit. State $3/2+$, which according to the theorem
given in Section IV cannot form a Wigner phase lies higher in
energy, like for the unperturbed dot.

The influence of the central attractive defect is qualitatively
different for odd and even electron number. For an odd number of
electrons it simply enhances the localization of the central
electron, and does not influence the ground-state degeneracy.
While for even $N$ it destroys Wigner crystallization leading to
the appearance of an extra charge maximum corresponding to
sub-electron charge and allows more states to become degenerate
with the ground-state.

\begin{figure*}[htbp]{\epsfxsize=130mm
                \epsfbox[33 447 582 642]{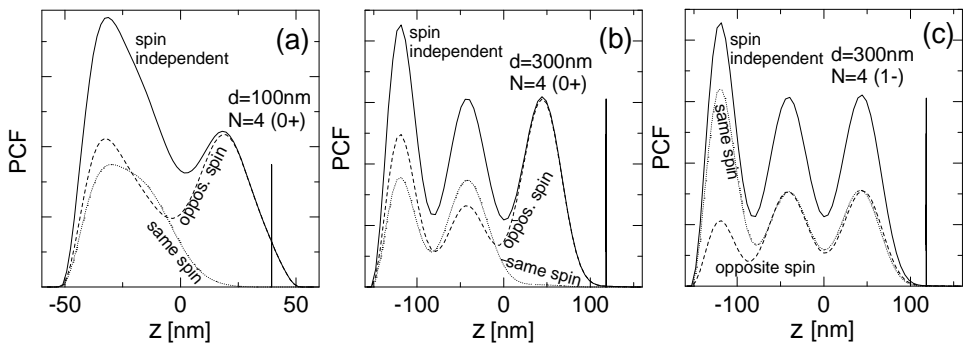}}\newline
\caption{Pair correlation functions (PCFs) for 4 electrons in
state $0+$ (a,b) and state $1-$ with $S_z=\hbar$ (c) for $d=100$
(a) and 300 nm (b,c). One of the electrons is fixed and its
position is marked by a thin vertical line. Solid curves show the
spin-independent PCF, dashed (dotted) curves show the opposite
(same) spin PCF.
 \label{spiny}
 }
\end{figure*}

\section{Spatial spin ordering in the Wigner limit}

It is interesting to look whether the low-spin ground states
exhibit any spatial antiferromagnetic ordering of the electron
spins. For even number of electrons and arbitrary dot length the
ground-state corresponds to zero total spin. In this case the
spin-up and spin-down densities are exactly equal to each other,
so that spin-ordering is not visible in the laboratory frame of
reference. In order to investigate a possible spin-ordering in the
zero-spin ground states one has to look into the inner coordinates
of the quantum system. We use here the spin-dependent pair
correlation functions (PCFs) defined for a given state by the
expectation values
\begin{widetext}
\begin{eqnarray}
F_{PCF}^{same}(z_a,z_b)=\left<\sum_{i=1}^N\sum_{j>i}^N
\delta(z_a-z_i)\delta(z_b-z_j)
\left(|\alpha(\sigma_i)\alpha(\sigma_j)><\alpha(\sigma_i)\alpha(\sigma_j)|+|\beta(\sigma_i)\beta(\sigma_j)><\beta(\sigma_i)\beta(\sigma_j)|\right)\right>,
\label{same}
\end{eqnarray}
and
\begin{eqnarray}
F_{PCF}^{oppo}(z_a,z_b)=\left<\sum_{i=1}^N\sum_{j>i}^N
\delta(z_a-z_i)\delta(z_b-z_j)
\left(|\alpha(\sigma_i)\beta(\sigma_j)><\alpha(\sigma_i)\beta(\sigma_j)|+|\beta(\sigma_i)\alpha(\sigma_j)><\beta(\sigma_i)\alpha(\sigma_j)|\right)\right>,
 \label{oppo}
\end{eqnarray}
\end{widetext}
where $\alpha$ and $\beta$ stand for spin-up and spin-down
eigenstates respectively. Functions (\ref{same},\ref{oppo}) give
the probability of finding at positions $z_a$ and $z_b$ a pair of
electrons with the same (\ref{same}) or opposite (\ref{oppo})
spins. The sum of functions ($\ref{same}$) and ($\ref{oppo}$)
gives the spin-independent PCF.

 Fig. \ref{spiny}(a) shows the PCF plots for
the 4-electron ground state in a small quantum dot [cf. Fig.
\ref{4e}(b)] with $d=100$ nm. The position of one of the electrons
is fixed near the right end of the dot [position marked by the
thin vertical line in Fig. \ref{spiny}(a)]. We see that the
probability of finding an electron with the same spin in the
neighborhood of the fixed-position electron is zero, which is a
signature of the Pauli exclusion principle. At the left side of
the dot probabilities of finding an electron with the same or
opposite spin as the one of the fixed position electron are nearly
equal. For the total zero-spin states in relatively small dots the
spin-ordering in the inner coordinates is of short range and
results from the Pauli exclusion. We only found a long-range
inner-coordinate spin-ordering in the Wigner crystallization
limit. Fig. \ref{spiny}(b) shows the plot for the 4-electron
ground state with $d=300$ nm. The charge density of the system
exhibits 4 distinct maxima [cf. Fig. \ref{4e}(d)]. We fix the
position of one of the electrons at the rightmost density maximum
[cf. the vertical line in Fig. \ref{spiny}(b)]. The probability
that the electron in the adjacent maximum has the opposite spin is
nearly 100 $\%$. The spin-dependent PCFs also differ for the two
charge maxima at the left of the origin. An electron confined at
the first (second) charge maximum to the left of the origin is
more probable to have the same (opposite) spin as the one of the
fixed electron. The ordering is of a probabilistic character, so
that the antiferromagnetic order of spins is the most probable to
be found, but the probability is not $100\%$. The spin-ordering in
this state has a clearly antiferromagnetic character and its range
covers the entire length of the dot. A similar inner-coordinate
antiferromagnetic order was previously found for quantum
rings.\cite{MKOSKINEN}

The $100\%$ probability of finding the opposite spin in the charge
maximum adjacent to the maximum associated with the fixed electron
presented in Fig. \ref{spiny}(b) is not, as one could naively
expect, related to the Pauli exclusion. In Fig. \ref{spiny}(c) we
plotted the PCF for the $1-$ state, which becomes degenerate with
the ground $0+$ state in the weak confinement limit. We see that
in this state the spin of electrons confined in the two central
maxima is independent of the spin of the electron at the rightmost
maximum. However, in this state one may expect that the electrons
at the opposite ends of the dot have the same spin, which means
that also in this state a long-range spin-ordering exists, even if
it is not of antiferromagnetic origin.

\begin{figure}[htbp]{\epsfxsize=75mm
                \epsfbox[100 180 520 635]{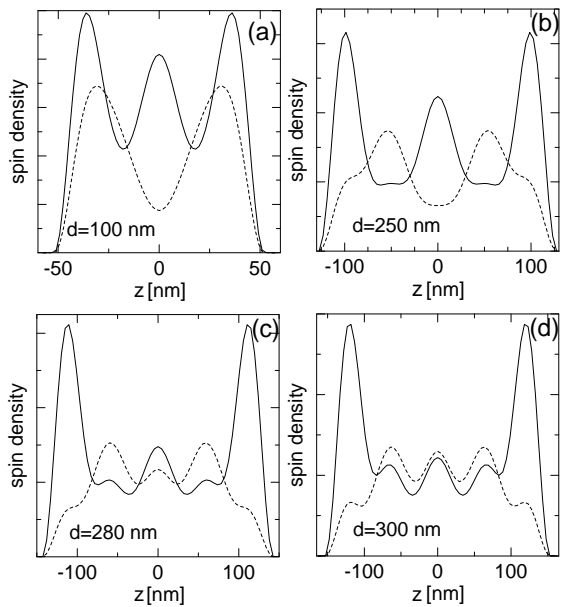}}\newline
\caption{ Spin-up (solid lines) and spin-down (dashed lines)
densities for the ground-state 5-electron system 1/2+ with
$S_z=\hbar/2$ for different system sizes.
 \label{5espin}
 }
\end{figure}

For odd number of electrons the difference between spin-up and
spin-down densities appears in the laboratory frame. This is
qualitatively different from quantum rings, which in fact are
endless structures. Fig. \ref{5espin}(a) shows the spin densities
for a relatively small dot length of $d=100$ nm [too small for the
ground-state Wigner localization to appear, cf. Fig. \ref{5e}(b)].
The spin-up electrons tend to gather at the extreme left and right
ends of the dot as well as in its center. The spin-down density is
minimal in the center of the dot, and the overall spin density
(difference of the spin-up and spin-down densities) exhibits
antiferromagnetic sign oscillations within the dot. These sign
oscillations are due to the electron-electron interaction since in
the noninteracting electron system the majority spin-up density is
nowhere smaller than the spin-down density. For larger systems
[$d=250$ nm, cf. Fig. \ref{5espin}(b)] the antiferromagnetic spin
oscillations become more pronounced. However, for even larger $d$
[cf. Fig. \ref{5espin}(c) and (d)], for which the Wigner molecule
appear in the 1/2+ ground state, the typically antiferromagnetic
real-space spin-ordering with the spin orientation changing
between the adjacent charge maxima vanishes.

\begin{figure*}[htbp]{\epsfxsize=130mm
                \epsfbox[55 453 564 637]{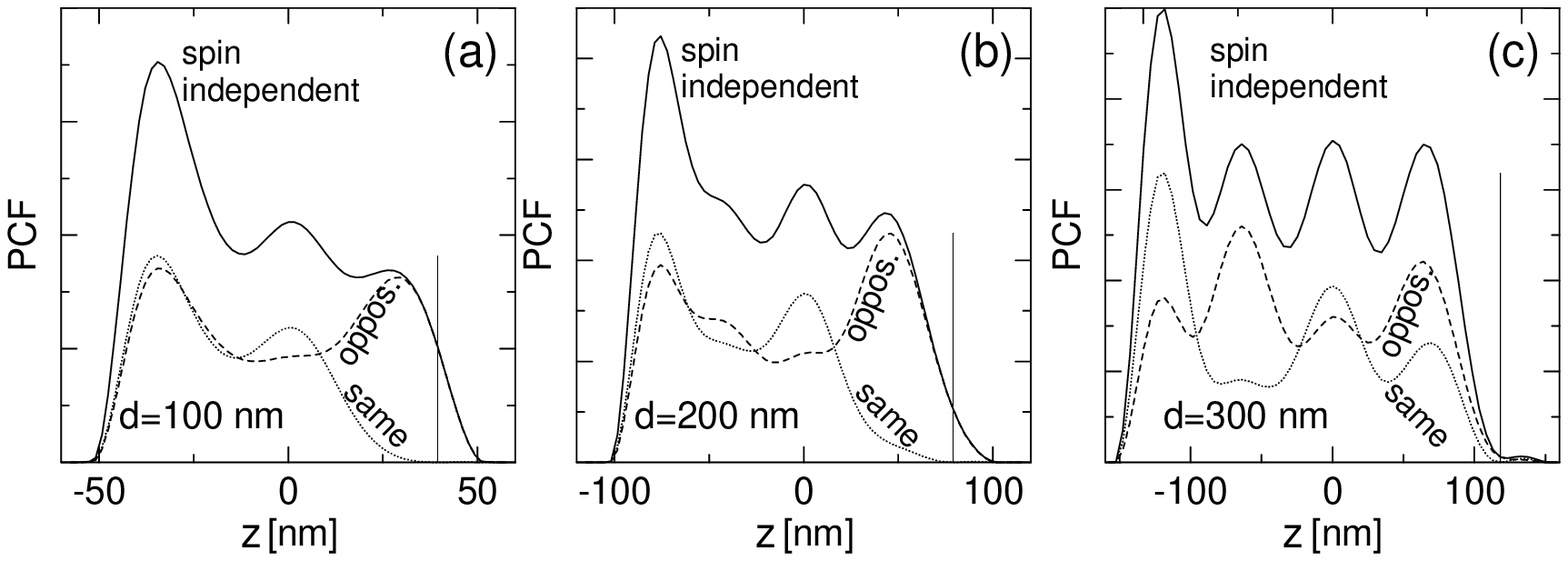}}\newline
\caption{ Pair correlation functions (PCFs) for 5 electrons in the
state $1/2+$ for $d=100$ (a), 200 nm (b) and 300 nm (c). The
position of the fixed electron is marked by a thin vertical line.
Solid lines show the spin-independent PCF, dashed (dotted) lines
show the opposite (same) spin PCF.
 \label{5epcf}
 }
\end{figure*}

Let us look at the spin distribution in the inner coordinates of
the 5-electron $1/2+$ ground state. Fig. \ref{5epcf}(a) shows the
PCF plots for $d=100$ nm. Electrons of the same spin as the fixed
electron do not appear in its close neighborhood, but are more
probable to be found at the center of the dot than electrons of
opposite spin. Probability of finding an electron at the opposite
side of the dot is independent of its spin. The spin order in this
relatively small dot ($d=100$ nm) is clearly short-range which is
similar as for the case of 4 electrons in a small dot [cf. Fig.
\ref{spiny}(a)]. The PCF plots for opposite spins at the left end
of the dot start to differentiate for $d=200$ nm [cf. Fig.
\ref{spiny}(b)]. For $d=300$ nm, for which Wigner localization is
observed [cf. Fig. 4(d)], the PCF plots show a long-range
antiferromagnetic spin ordering. Notice the growth of the PCF plot
for the same spin direction in the closest neighborhood of the
fixed-position electron from $d=200$ to 300 nm in Figs.
\ref{spiny}(b) and (c). Pauli exclusion plays a less significant
role for larger distances between the charge maxima.

Density-functional studies\cite{Nieminen,Reiman} predict the
appearance of interlocked waves of opposite spins in the
laboratory frame for long quasi one-dimensional dots. The
appearance of the spin-density wave for even electron number
amounts in spin symmetry breaking. Recently,\cite{Nieminen} it was
found that for even $N$ the formation of the spin density wave in
the density functional theory accompanies the Wigner
crystallization. But in the present study we find that for the
exact solution spin symmetry is conserved and Wigner
crystallization is associated with the inner space spin ordering.
In the exact solution the interlocked spin densities in the
laboratory frame can only be observed for odd numbers of
electrons, but the presented 5-electron case shows that this
effect is not necessarily related with Wigner crystallization. In
the exact solution the electrons with opposite spins avoid one
another in the inner space. A mean field approach can only account
for this effect by symmetry breaking. The reason for the
occurrence of spin symmetry breaking in the mean field approach
for large single-dimensional dots are similar to the origin of the
broken spatial symmetry mean field solutions for the magnetic
field induced Wigner crystallization in circular
structures.\cite{ReiMan}

In large systems the spin-independent PCF plots become identical
for all states degenerate with the ground-state [cf.
spin-independent PCFs for the 4-electron degenerate $0+$ and $1-$
states in Figs. \ref{spiny}(b) and (c)]. This means, that in
Wigner-molecule states electrons avoid one another with the same
efficiency independently of their spins. As a matter of fact this
is the origin of the appearance of the ground-state degeneracy in
the Wigner molecule regime. One-dimensional Wigner molecules
present pronounced magnetic properties related to the long-range
spin-ordering in the inner coordinate space. This ordering for
different degenerate spin eigenstates may be typical for
ferromagnetic, antiferromagnetic or even an other form of order.
Due to the vanishing energy spacing between the different spin
states the spin magnetic properties of Wigner molecules are of a
very soft character. The Wigner molecules should be extremely
susceptible to any spin-dependent interactions. In particular even
a weak additional effect promoting the spin-polarized phase can
result in spin-polarization of the system. A possible
spin-polarization of the one-dimensional electron gas has been
found\cite{PSP} in transport measurements.

\section{conclusions and summary}

We have studied the ground and excited states of electron systems
confined in quasi-one-dimensional quantum dots using an exact
diagonalization approach. For large systems we found Wigner
localization which appears not only in the ground state but also
for several excited states which eventually leads to the
degeneracy of the ground-state in the large $d$ limit. We have
considered spin and spatial parity of states forming Wigner
molecules. We have shown that the parity of the spin-polarized
state which forms a Wigner molecule is strictly determined by the
number of electrons.

We have discussed the effect of a central attractive defect which
destroys Wigner crystallization for an even number of electrons
allowing more states to become degenerate with the ground-state in
the weak confinement limit. For odd electron numbers the central
defect enhances the localization of the electron occupying the
central position in the Wigner molecule and does not affect the
ground-state degeneracy.

We have investigated the spin ordering effects associated with
Wigner crystallization. We have found that for small dots the
spatial spin-ordering in the inner coordinates has a short-range
character and results mainly from the Pauli exclusion principle.
The long-range spatial spin order appears only in the Wigner
molecule regime when the electrons occupy distinct sites within
the quantum dot. We conclude that in one-dimensional quantum dots
the Wigner crystallization is a necessary condition for the long
range spin ordering to appear. We have identified the effect of
spin symmetry breaking observed in the density functional theory
as a tendency of the mean field method to mimic the internal-space
spin ordering present in the exact solution for the Wigner
molecule regime.

{\bf Acknowledgments} This paper has been partly supported by the
Polish Ministry of Scientific Research and Information Technology
in the framework of the solicited grant PBZ-MIN-008/P03/2003, by
the Flemish Science Foundation (FWO-Vl), the Belgian Science
Policy and the University of Antwerpen (VIS and GOA). B. Szafran
is supported by the Foundation for Polish Science (FNP) and T.
Chwiej is partially supported by a Marie Curie training fellowship
of the European Union.
\newline


\begin{thebibliography}{00}
\bibitem{2doN} The approximate ground-state degeneracy of a $N$ electron
system in large one-dimensional dot equals $2^N$ (taking into
account the degeneracies with respect to the $z$-component of the
total spin) as reported already in Ref. [\cite{Hkramer}]. For 4
and 5 electrons also higher-energy states for a given spin and
parity quantum numbers (not discussed in the present paper) become
degenerate with the ground state. We have determined the parity of
these states. Namely, the 4-electron ground state becomes
degenerate with second $1-$ and $0+$ states and the 5-electron
ground state becomes degenerate with the second $1/2-,3/2-,1/2+$
and $3/2+$ symmetry states as well as with the third $1/2+$ state.
\bibitem{1sd} A single
determinant is sufficient to describe the spin-polarized Wigner
molecule formed by $N$ electrons localizated around $N$ space
sites. See also the discussion of the electron-electron
correlation given in Ref. [\cite{2e}].
\end{thebibliography}
\end{document}